\documentclass[12pt]{article}
\usepackage{graphicx}

\newcommand\pubnumber{DPF2015-249}
\newcommand\pubdate{\today}

\def\napoli{Perimeter Institute for Theoretical Physics,
31 Caroline St. N, Waterloo, Ontario, Canada N2L 2Y5.}
\def\support{\footnote{Work supported by the Government of Canada through Industry Canada and by
the Province of Ontario through the Ministry of Economic Development $\&$ Innovation.}}

\def\Title#1{\begin{center} {\Large #1 } \end{center}}
\def\Author#1{\begin{center}{ \sc #1} \end{center}}
\def\Address#1{\begin{center}{ \it #1} \end{center}}

\newcommand\pubblock{\rightline{\begin{tabular}{l} \pubnumber\\
         \pubdate  \end{tabular}}}
\newenvironment{Abstract}{\begin{quotation}  }{\end{quotation}}
\newenvironment{Presented}{\begin{quotation} \begin{center} 
             PRESENTED AT\end{center}\bigskip 
      \begin{center}\begin{large}}{\end{large}\end{center} \end{quotation}}
\def\Acknowledgments{\bigskip  \bigskip \begin{center} \begin{large}
             \bf ACKNOWLEDGMENTS \end{large}\end{center}}

\def\beq{\begin{equation}}
\def\eeq#1{\label{#1}\end{equation}}
\def\eeqn{\end{equation}}

\def\beqa{\begin{eqnarray}}
\def\eeqa#1{\label{#1}\end{eqnarray}}
\def\eeqan{\end{eqnarray}}

\let\bar=\overbar

\def\Dslash{\not{\hbox{\kern-4pt $D$}}}
\def\dslash{\not{\hbox{\kern-2pt $\del$}}}

\def\msb{{\bar{\ssstyle M \kern -1pt S}}}

\begin{document}
\begin{titlepage}
\pubblock

\vfill
\Title{Exploration of Beyond the Standard Model Physics at the ILC}
\vfill
\Author{ Stefania Gori\support}
\Address{\napoli}
\vfill
\begin{Abstract}
The International Linear Collider (ILC) will advance our understanding of fundamental physics, through a program for precision measurements of the Higgs boson and of the top quark properties. An additional crucial goal of the ILC will be the direct discovery of light New Physics (NP) particles. 
This proceeding gives an overview of the ILC direct discovery potential. 
Special emphasis is dedicated to the discovery prospects for light hidden Supersymmetric particles, such as Higgsinos and staus.  Furthermore, it will be extremely promising to have analyses aimed to the discovery of light hidden NP particles produced from new decay modes of the Higgs boson. 
The ILC low background environment, as well as the precise knowledge of the initial-state beams are also ideal for the precise measurements of the spectrum and interactions of these NP particles, once discovered.

\end{Abstract}

\vfill
\begin{Presented}
DPF 2015\\
The Meeting of the American Physical Society\\
Division of Particles and Fields\\
Ann Arbor, Michigan, August 4--8, 2015\\
\end{Presented}
\vfill
\end{titlepage}
\def\thefootnote{\fnsymbol{footnote}}
\setcounter{footnote}{0}
%

\section{Introduction}
With the LHC discovery of the 125 GeV Higgs boson, it is critically important to determine the properties of this new particle as comprehensively and as accurately as possible. The International Linear Collider (ILC) will cover a special role in measuring the various Higgs couplings with a precision at the (sub)percent level.

The ILC precision program will be complemented by a rich program of searches for direct production of New Physics (NP) particles. In particular, the ILC will have in several cases a unique sensitivity in searching for NP signals arising from the electroweak production of new particles, assuming these are kinematically accessible. This is especially important, in view of the null-results in the search for $SU(3)$ charged particles at Run I LHC. Very interesting examples are searches for light electroweak Supersymmetric (SUSY) particles, like Higgsinos and staus, as well as searches for exotic decays of the 125 GeV Higgs boson into light singlet-state particles.

At Run II LHC (LHC14), the production cross section of a pair of Higgsinos and a pair of staus is relatively sizable: $\sim 300$ fb for 200 GeV charged Higgsinos and $\sim 20$ fb for 200 GeV staus with an equal left and right-handed component. Nevertheless, these particles are quite challenging for the LHC, due to their hidden signatures, leading to background limited LHC searches.
Higgsinos and staus may well be observable in a
clean environment as at the ILC, as long as they are kinematically accessible (Sec.~\ref{sec:lightSusy}).

Furthermore, in addition to the precise measurement of the Standard Model (SM) Higgs couplings, the determination of the eventual Higgs couplings to light ($m_h>M_{\rm{NP}}/2$) NP particles through Higgs exotic signatures represents a crucial physics program to unveil physics beyond the Standard Model (BSM) and, in particular, to test hidden sectors (see~\cite{Curtin:2013fra} for a review). In Sec.~\ref{sec:DMandHiggs}, we will discuss the complementarity between possible LHC and ILC measurements of Higgs exotic decays.

\section{Light Hidden Supersymmetry}\label{sec:lightSusy}

In Supersymmetric extensions of the Standard Model, Higgsinos are preferred to have masses at around the electroweak scale by naturalness arguments, since the $\mu$ parameter enters the tree-level Higgs potential. In SUSY spectra with Higgsinos Lightest Supersymmetric Particle (LSP) and with multi-TeV gaugino masses, the
mass splittings between charged and neutral Higgsino-like states are relatively small and of the order of 1-5 GeV. For much heavier gaugino masses, radiative corrections to the Higgsino-like chargino and neutralino masses are relevant and lead to a splitting of the order of $\sim 200$ MeV. 

At colliders, Higgsino production proceeds through the s-channel exchange of a $Z$ boson, a photon or a $W$ boson: $pp(e^+e^-)\to\tilde \chi_2^0\tilde\chi_2^0,\tilde\chi_1^\pm\tilde\chi_1^\mp,\tilde\chi_2^0\tilde\chi_1^\pm$, followed by the decay of $\tilde\chi_2^0$ and $\tilde\chi_1^\pm$ into (soft) visible particles and missing energy, carried in great part by the LSP $\tilde\chi_1^0$. This type of spectra is notoriously challenging for the LHC due to the small amount of missing energy and to the low $p_T$ spectrum of the visible particles. One known way to improve the search sensitivity to small mass-gap regions is to
utilize a relatively hard initial state (ISR) jet, that results in an overall boost of the $\tilde\chi_2^0\tilde\chi_2^0,\tilde\chi_1^\pm\tilde\chi_1^\mp$ and $\tilde\chi_2^0\tilde\chi_1^\pm$ systems. Nevertheless, even using this technique, LHC analyses based on Run I data could not push the reach on Higgsinos significantly beyond the LEP reach, if the mass splitting is below (20-30) GeV~\cite{Aad:2015eda}.

\begin{figure}[thb]
\centering
\includegraphics[height=2.0in]{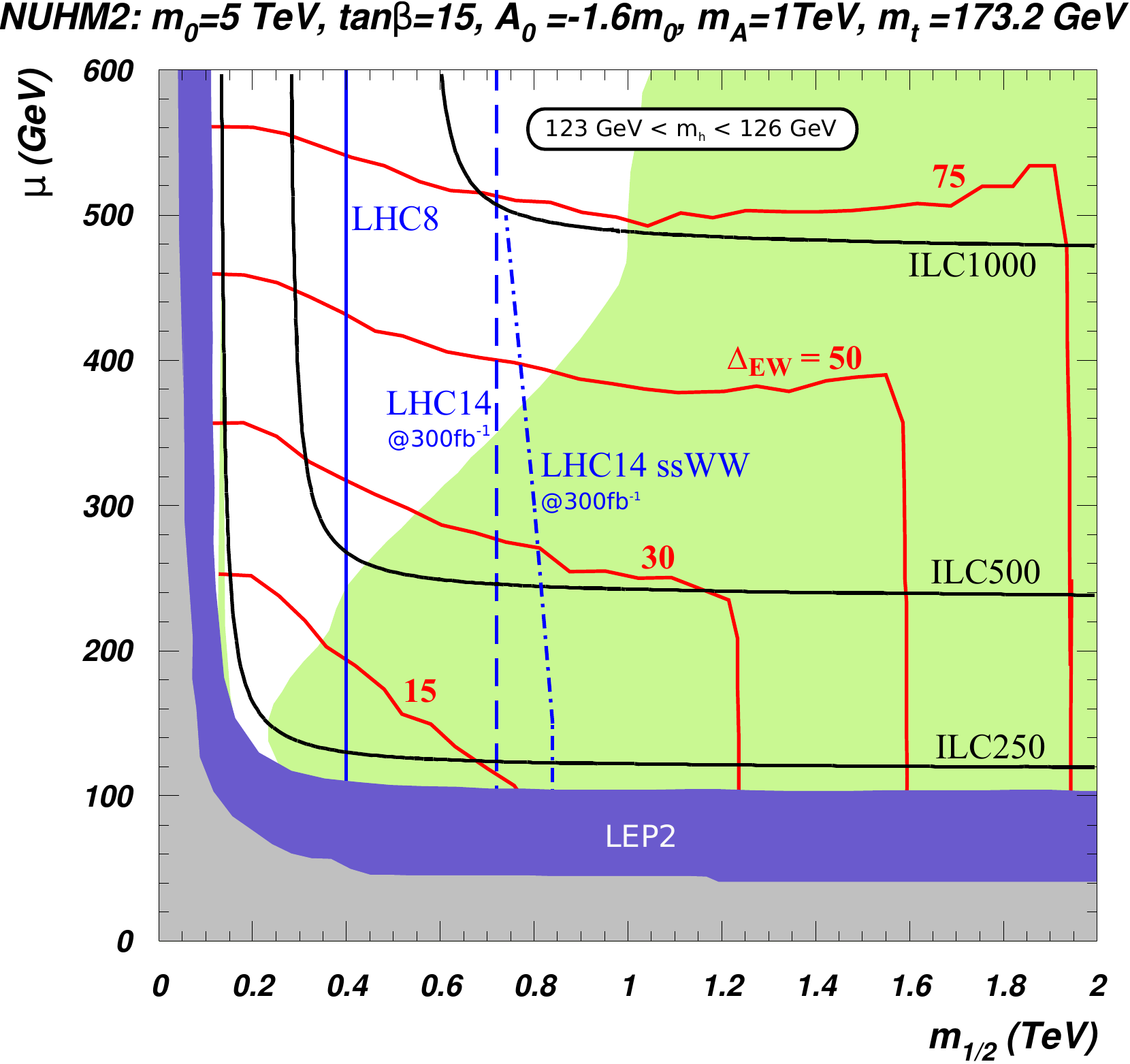}
\includegraphics[height=2.0in]{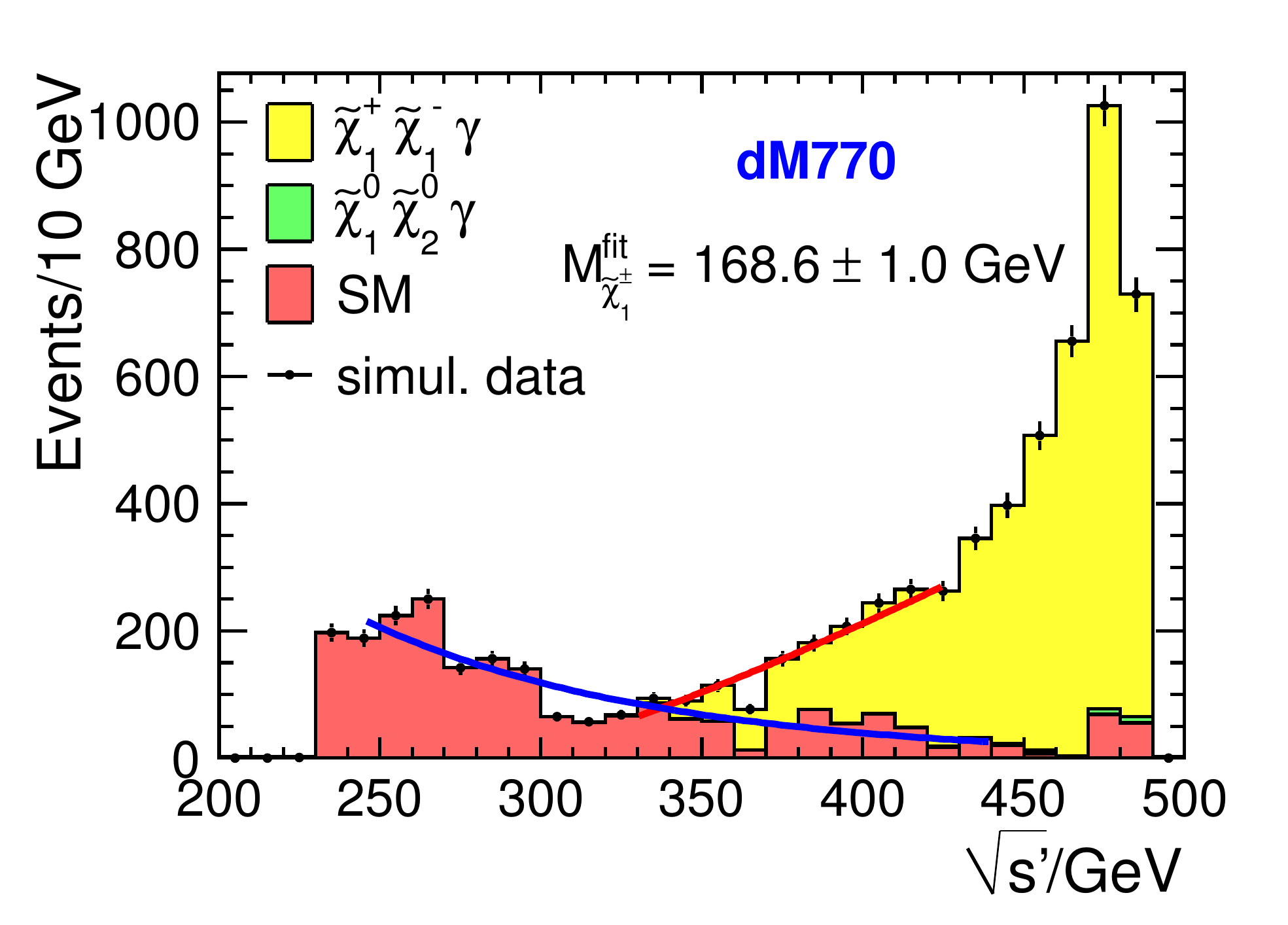}
\caption{Left panel (from~\cite{Baer:2013faa}): ILC discovery reach for Higgsino pair production (black contours). Also shown the reach for gluino pair production at Run I LHC 
(solid blue contour), and the region accessible to LHC14 searches of gluino cascade decays with 300 fb$^{-1}$ data (dashed and dot-dashed blue contours). Right panel (from~\cite{Berggren:2013vfa}): distribution of the reduced center-of-mass energy of the system recoiling against the ISR photon for the benchmark {\it{dM770}} with $m_{\tilde\chi_1^\pm} = 167.36$ GeV, $m_{\tilde\chi_1^0} = 166.59$ GeV, $m_{\tilde\chi_2^0}= 167.63$ GeV.}
\label{fig:Higgsino}
\end{figure}

At the ILC, thanks to the precisely known initial state, an ISR photon can be used very efficiently to test squeezed Higgsino scenarios\footnote{Note that this applies generically to squeezed electroweak spectra.}. This results in a possible quick discovery of Higgsinos once they are kinematically accessible (see the left panel of Fig.~\ref{fig:Higgsino} for the discovery reach in the $\mu-m_{1/2}$ plane).

Furthermore, the ILC has the opportunity to precisely determine the masses of the Higgsinos, once discovered. The chargino mass can be reconstructed by fitting the distribution for the reduced center-of-mass energy $\sqrt{s^\prime}$ ($s^\prime=s-2\sqrt s E_\gamma$) near the endpoint region (see the red line in the right panel of Fig.~\ref{fig:Higgsino}). This gives a determination of the chargino mass at the level of $\mathcal O(1\%)$. Additionally, the mass splitting between the chargino and the LSP can be determined by fitting the energy distribution of the pion arising from the Higgsino decay with an uncertainty at the level of $\mathcal O(5\%)$~\cite{Berggren:2013vfa}.

Higgsinos are not the only SUSY particles that can be hidden to the LHC. Light staus represent another theoretically well motivated example, since they arise e.g. in gauge-mediated supersymmetry breaking models, as well as in Dark Matter (DM) models with stau-DM co-annihilation. At the LHC, direct searches for promptly decaying staus are notoriously difficult (see for example~\cite{Carena:2012gp}), due to 
the challenge in disentangling a di-tau plus missing energy signature ($pp\to\tilde\tau_1\tilde\tau_1\to 2\tau+$MET) from the W+jets and di-boson SM backgrounds. Present LHC searches based on Run I data could not push the reach beyond the one obtained by LEP~\cite{Aad:2015eda}.

\begin{figure}[thb]
\centering
\includegraphics[height=2.0in]{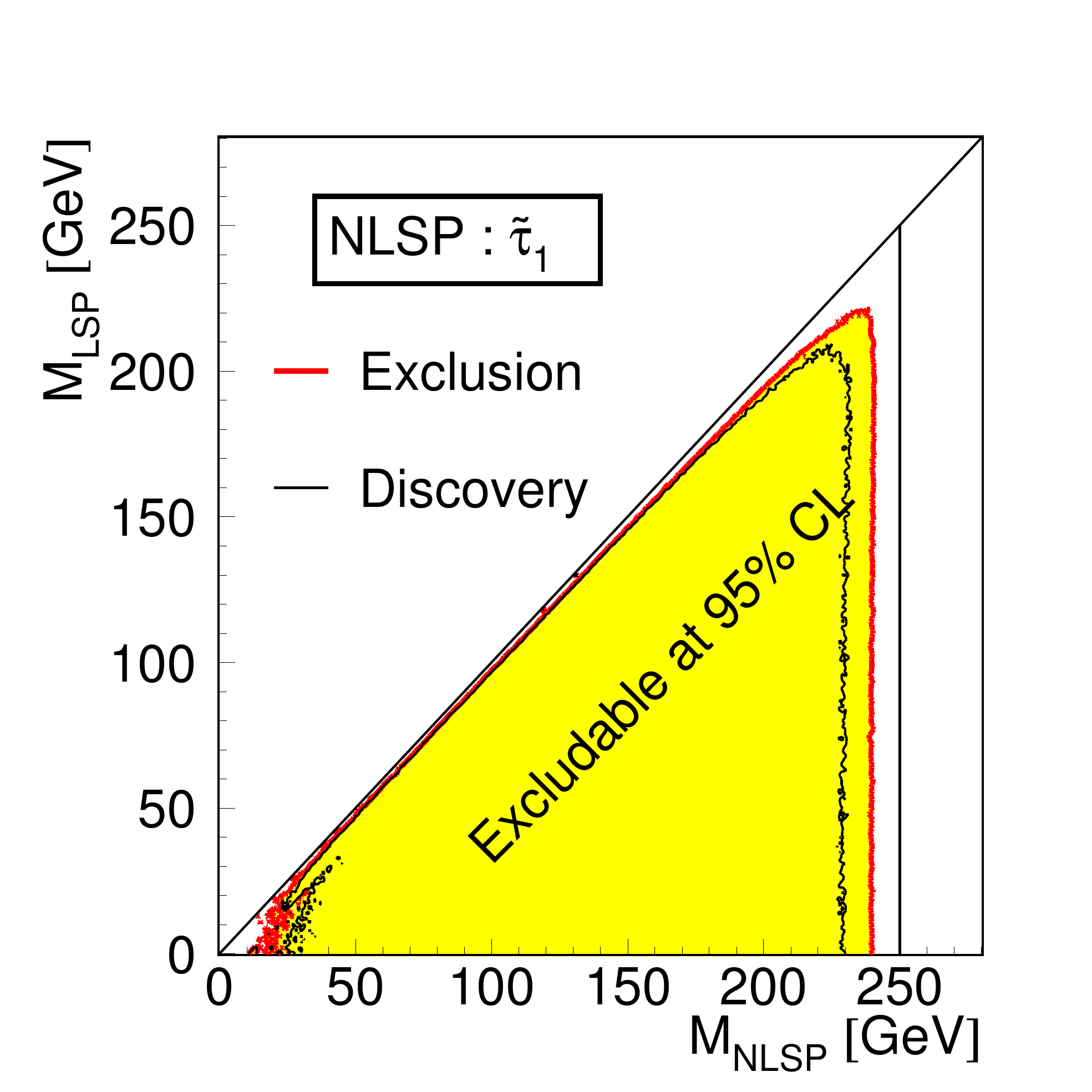}
\includegraphics[height=2.0in]{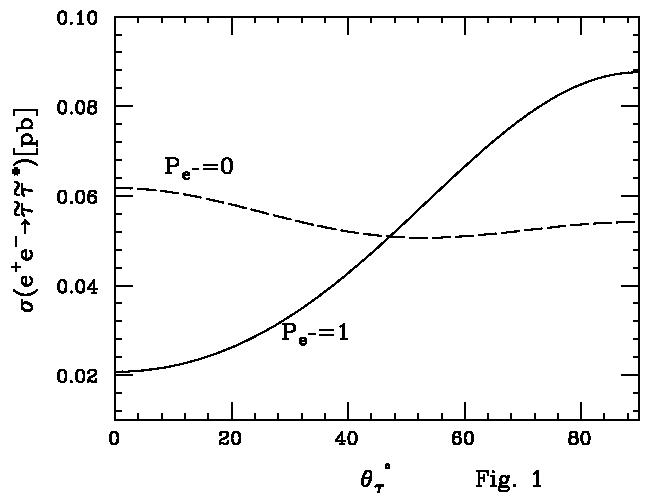}
\caption{Left panel (from~\cite{Baer:2013vqa}): exclusion bound (in red) and discovery reach (in black) for a $\tilde\tau_1$ NLSP with 500 fb$^{-1}$ at $\sqrt s=500$ GeV. Right panel (from~\cite{Nojiri:1994it}): Cross section for $\tilde\tau_1\tilde\tau_1$ production as a function of the stau mixing angle $\theta_\tau$, for two different electron polarizations $P_{e^-}$.}
\label{fig:Staus}
\end{figure}

Due to the cleaner ILC environment, the ILC could probe staus up to masses very close to the kinematical reach, if the mass splitting between the stau and the LSP is larger than $\sim 20$ GeV (see the red contour in the left panel of Fig.~\ref{fig:Staus}). Additionally, also squeezed spectra with $m_{\tilde\tau_1}-m_{\rm{LSP}}=\mathcal O({\rm{few \,GeV}})$ could be probed for stau masses up to $\sim 200$ GeV, at $\sqrt s=500$ GeV. Once a stau is discovered, the ILC will be able to perform a large set of precision measurements of the stau spectrum. In particular, the stau mass can be determined using threshold scans with an uncertainty at the level of $\mathcal O(1)$ GeV (see for example~\cite{Belanger:2008yc}). Finally, one can measure the left-right mixing of $\tilde\tau_1$ and the Higgsino/gaugino mixing of the LSP 
by measuring both the cross section of $\tilde\tau_1$ production (see right panel of Fig.~\ref{fig:Staus} for the dependence of the cross section on the stau mixing angle $\theta_\tau$) and the average polarization of the $\tau$ from the $\tilde\tau_1$ decay~\cite{Nojiri:1994it}.

\section{Dark Matter and Higgs}\label{sec:DMandHiggs}
Dark Matter dominates the matter density in our Universe, but very little is known about it. One of the prime tasks of the LHC and of future colliders is to identify the nature of Dark Matter. 
WIMP Dark Matter can be searched for by looking for mono-jet or mono-photon events. These searches are typically interpreted in terms of bounds on higher dimensional effective operators mediating the interactions of DM with SM particles and suppressed by a NP scale $\Lambda$. For WIMP masses below
the kinematic limit, the ILC reach in the effective operator scale is rather independent of the WIMP mass and is $\Lambda\sim 3$ TeV in the case of vector (D5) and axial-vector (D8) operators~\cite{Barklow:2015tja} (see also~\cite{Chae:2012bq}).

Beyond the canonical WIMPs, also alternative models for DM have been developed. In particular, DM can belong to a hidden sector and carry its thermalization thanks to its interactions with BSM particles, not charged under the SM gauge group. Such particles could easily have escaped past
experimental searches, since they interact only feebly with ordinary matter. 

The Higgs boson represents a unique opportunity to test hidden sectors, as it can serve as a portal, being the $HH^\dagger$ operator the lower dimensional singlet that can be written out of SM fields. As an example, several well motivated BSM models contain a hidden sector scalar that couples to the Higgs through the Higgs portal operator $ \xi |H|^2|S|^2$. If the scalar is lighter than $m_h/2$, this operator will generically lead to Higgs exotic decays of the type $h\to S S$ and therefore to a variety of new Higgs signatures, depending on the decay mode of the scalar $S$. 

These exotic decays can be looked for at the ILC in a twofold way: either indirectly, through a precise measurement of the total Higgs width, or directly, looking for the new Higgs decay modes. The former method will be able to unravel the existence of new Higgs decay modes with branching ratios at the level of 3.8$\%$ at an Initial Phase and of 1.8$\%$ with the full data set\footnote{As presented in~\cite{Fujii:2015jha}, the Initial Phase will have 500 fb$^{-1}$ at 500 GeV, 200 fb$^{-1}$ at 350 GeV, and 500 fb$^{-1}$ at 250 GeV. The luminosity-upgraded phase will have an additional 3500 fb$^{-1}$ at 500 GeV and 1500 fb$^{-1}$ at 250 GeV.}, 
using the recoil technique that allows the measurement of the ZH cross section without reconstructing the decays of the Higgs
boson~\cite{Fujii:2015jha}. 
Additionally, present and future colliders like the ILC can look directly for exotic decays of the Higgs boson. The LHC will produce $\sim 1.5\times 10^7$ Higgs bosons ($\sim 1.5\times 10^8$ at the High-Luminosity (HL) stage), offering a unique opportunity to probe tiny branching ratios for clean decay modes, as for example for the Higgs decay to dark photons ($Z_D$), $h\to Z_DZ_D\to 4l$, for which the reach is at the level of $\mathcal O(10^{-7})$ at the HL-LHC~\cite{Curtin:2014cca}.

On the contrary, the ILC will produce $\sim 3\times 10^5$ Higgs bosons during the Initial Phase and $\sim 2 \times 10^6$ Higgs bosons with the full data set, potentially probing branching ratios into NP particles as small as $\mathcal O({\rm{few}}\times 10^{-6})$. 
In spite of the smaller number of Higgs bosons produced, the ILC can play a significant role in
searching for those exotic decays that are more challenging for the LHC, because background limited. Typical examples are 
$h\to SS\to 4\tau$ and $h\to SS\to 4b$, that arise naturally in many theories beyond the Standard Model, as for example in the Next to Minimal Supersymmetric Standard Model (NMSSM). In the case of $h\to SS\to 4\tau$, Ref.~\cite{Liu:2013gea} shows a possible reach at the level of $\mathcal O({\rm{few}}\times 10^{-3})$. It will be very interesting to perform a dedicated study to determine the ILC reach for $h\to SS\to 4b$.

\section{Conclusions}
An $e^+e^-$ collider like the ILC offers a unique opportunity to discover new phenomena indirectly, through the precise measurements of Higgs and top couplings.
At the same time, as we have stressed in this talk, the ILC 
is a machine
for direct discoveries. In particular, the ILC will be able to test $SU(3)$-singlet new physics particles that are hidden to the LHC, but nevertheless extremely well motivated, such as Higgsinos and staus, if they are within the kinematical reach. Furthermore, to have a complete program for Higgs characterization, it will be extremely promising to have ILC analyses aimed to the discovery of exotic Higgs decay modes into light new physics particles, that are otherwise hidden to hadron
colliders.

\Acknowledgments
I am grateful to Michael Peskin for discussions.

\end{document}